\documentclass[aip,jmp,bmf, sd, rsi, amsmath,amssymb, reprint]{revtex4-1}
\usepackage{graphicx}
\usepackage{bigints}

\begin{document}

\title{Nanoscale ballistic diodes made of polar materials for amplification and generation of radiation in 10 THz-range}
\author{V. A. Kochelap}
\author{V. V. Korotyeyev}
\email[E-mail:]{koroteev@ukr.net}
\author{Yu. M. Lyashchuk}
\address{Department of Theoretical Physics, Institute of Semiconductor Physics, Kyiv 03650, Ukraine}

\author{K. W. Kim}
\address{Department of Electrical and Computer Engineering,
North Carolina State University, Raleigh, North Carolina 27695-7911}

\begin{abstract}
We investigate ultra-high frequency electrical properties of
nanoscale $n^+ - i -n^+$ diodes made of polar semiconductors. The
calculations show that the coupling between optical vibrations of
the lattice and the ballistic electrons strongly modifies and
enhances the time-of-flight effects giving rise to narrow
resonances of the diode impedance in the reststrahlen frequency range.
Particularly, negative dynamic resistance is induced in close proximity
to  the optical phonon frequency. The resonant effects in the dynamic resistance
of nanoscale GaAs and InP diodes are studied in detail. The obtained magnitudes of
the negative dynamic resistance effect indicate that the nanoscale diodes are capable
of generating electromagnetic radiation in far-infrared spectral range under electric pumping.

\end{abstract}
\maketitle

\section{Introduction}

The interaction of ballistic electrons with the electromagnetic
radiation offers a significant opportunity for generation and
detection of ultra-high frequency radiation. Particularly, the two-
and three-terminal structures scaled down to the nanometer
dimensions can access the THz frequency range by taking advantage
of the ultra-fast semiclassical {\em ballistic electron
transport}.~\cite{Dyak,TWT-1,Ryzhii-1,Grib,Grib-1} It is believed that
nanoscale ballistic devices have already reached to the point
where their basic effects and operational principles are very
similar to those of the electron vacuum tubes exploiting the
time-of-flight effects (see, for example, Ref~\onlinecite{Tubes}).
However, semiconductor materials possess a number of inherent
properties different from the vacuum tubes. For example, the
nonparabolic energy dispersion in the semiconductors  can lead to
the negative effective masses~\cite{Grib} and the velocity
saturation~\cite{Grib-1} at large electron energies,  which drastically
modify both steady state and high frequency parameters of
the ballistic diodes.

Another important factor giving rise to characteristic features of
solid-state ballistic devices is the frequency dependence of the
dielectric permittivity $\kappa (\omega)$. Indeed, the high
frequency response of these devices and, particularly, the
negative dynamic resistance (NDR) effect arise due to the
correlation between the electronic motion and the space charge
dynamics.~\cite{Ryzhii-1,Grib,Grib-1,JNO, Firsov1}  The latter are strongly
influenced by the screening effects. Polarization of the crystal
lattice also contributes to the screening. In the important case
of {\em polar} materials, the lattice polarization via optical
vibrations determines the dielectric permittivity as:~\cite{Born}
\begin{equation} \label{dielectric}
\kappa (\omega) =  \kappa_{\infty} + \frac{(\kappa_0 -
\kappa_{\infty})\, \omega_{TO}^2}{\omega_{TO}^2 - \omega^2 - 2 i
\gamma \, \omega},\end{equation} where $\kappa_{0}$ and
$\kappa_{\infty}$ are the low frequency and high frequency
permittivities, respectively, $\omega_{TO}$ is the frequency of the
transverse optical vibrations, and $\gamma$ is the optical phonon
damping. The frequency of the longitudinal optical vibrations,
$\omega_{LO}$, equals to $\sqrt{\frac{\kappa_{0}}{\kappa_{\infty}}}
\omega_{TO}$. The permittivity, $\kappa (\omega)$,
varies considerably near the frequency $\omega_{LO}$ and $Re[\kappa(\omega)]$
can change its sign in the interval $[\omega_{TO}, \omega_{LO}]$,
which is known as the {\em reststrahlen frequency range} (RFR).~\cite{Cardona,Adachi}

When in a diode, the electron time-of-flight, $\tau_{tr}$,
is of the order of $2\pi/\omega_{LO} $, one can expect resonant
effects in high frequency resistivity. Physics-based explanation
of expected effects is obvious. Under the mentioned condition,
characteristic frequencies of ballistic charge transfer across
the diode are in the resonance with the polarization lattice vibrations.
This results in a sharp enhancement of mutual influence of the electron
and polarization dynamics in the narrow RFR.
In general, a time dependent perturbation of the voltage applied to the diode
can be presented as a Fourier expansion with different frequencies. One can suppose
that the Fourier components of the diode response (the current) to this perturbation
with frequencies close to $\omega_{LO}$ will be resonantly increased.

Note, for semiconductor materials and heterostructures widely used in high
speed electronics, the characteristic frequencies of optical phonons
vary in a wide THz range. In III-V compounds $\omega_{LO}$ are in the range
5 THz (InSb) to 12 THz (GaP)~\cite{Levinshtein};
for the group-III-nitrides $\omega_{LO}$ are of 18 to 22 THz~\cite{Adachi, Firsov2},
etc.  Currently, for this frequency range the physical effects are actively
studied,~\cite{Leit-1,Leit-2,Tsen-1,Jap,PRL-2019}
as well as their device applications.~\cite{Gelmont,Kim,Shur,Kukhtaruk,Sydoruk,Ohtani,Khurgin-1,polaron-l-2}

In this paper, we investigate ultra-high frequency electron
response of nanoscale ballistic diodes. For the nanoscale
structures made of a polar material, it is found that the polarization
lattice vibrations provide the dynamic screening and contribute
to the resonant high frequency response, leading
to large effects in the resistivity and, particularly, to an enhanced NDR
effect in the RFR. This mechanism can provide an electrical
means to amplify and generate electromagnetic radiation in 10--THz
frequency range, i.e., far-infrared radiation.

To study the resonant enhanced effects in the resistivity
for the RFR, we analyze a short $n^+ - i - n^+$ diode assuming ballistic
electron motion and frequency dependent permittivity of
Eq.~(\ref{dielectric}) in the $i$-region ($0<x<L$).
Two models of the electron transport are considered:
the model of monoenergetic electron injection into the {\it i}--region (the base) and
the model based on the Boltzmann transport equation for
 the injected electrons.

\section{Model of monoenergetic electron injection} \label{S-2}

We begin with the analysis of a simple model of the diode with injection of monoenergetic
electrons from cathode to the $i-$base. This model facilitates the understanding of main features
of the high-frequency electron dynamics and resistivity of the diode in the RFR.
To describe the semiclassical space-charge-limited
transport of ballistic electrons we introduce the velocity, $V(x,t)$, the electron density,
$n(x,t)$, the electric field, $F(x,t)$, and the electron current
density, $J (x,t) = - e n(x, t)V (x, t)$. Here the $x$ coordinate
varies along the diode base: $0\leq x \leq L $.
For this approach, the basic equations are the
Newton's law, the continuity equation, and the Poisson
equation:\noindent
\begin{eqnarray} \label{N}
\frac{\partial V}{\partial t} + V \frac{\partial V}{\partial x} = - \frac{e}{m} F,\\
\frac{\partial J}{\partial x} - e \frac{\partial n}{\partial t} =0,
\\ \label{C}\frac{\partial D}{\partial x} = - 4 \pi e n \label{P}\,.
\end{eqnarray}
Here, $m$ is the electron effective mass and $D(x,t)$ is the electric
displacement.
In Eqs.~(\ref{N})-(\ref{P}), every variable can be presented as a sum of steady
state and time dependent contributions: $V = V_0 (x)+ V_{\omega} (x) \exp (- i \omega t),\,
F= F_0 (x) + F_{\omega} (x) \exp (- i \omega t)$, $J = J_0 (x)
+J_{\omega} (x) \exp (- i \omega t)$, etc. Then, one can define $D=D_0(x) + D_{\omega}(x) \exp (- i \omega t)$
with
$D_0 =\kappa_0 F_0$ and $D_{\omega} =\kappa(\omega) F_{\omega}$.
The expression for $\kappa (\omega)$ is as provided in Eq.~(1).
The dynamic lattice polarization, $P_{\omega} (x)$, can be
calculated via the alternative electric field $F_{\omega} (x)$:
\begin{equation} \label{PP}
P_{\omega} (x) = \frac{\kappa (\omega) - \kappa_{\infty}}{4 \pi} F_{\omega} (x)\,.
\end{equation}

 From Eqs.~(3) and (\ref{P}), it follows that the sum of the
conductivity current and the displacement current (i.e., the total
current) is constant throughout the diode. This leads to $J_{0}
=constant$ and $ J_{\omega} (x) - i  \omega \frac{ \kappa
(\omega)}{4\pi} F_{\omega} (x) = J_{\omega} = constant$ for the
steady state and time dependent problems, respectively.  With
given $J_0$ and $J_{\omega}$, both problems are reduced to
determining two unknown functions, for example, $V (x)$ and $F (x)$.

Below we use the "virtual cathode approximation''
\cite{virtual}, correspondingly, we set $V_0
(0) =V_i$ (i.e., the initial velocity) and $F_0 (0) = 0$ for the
steady state problem, and $V_{\omega} (0) =  F_{\omega} (0) =  0$ for
the time dependent case, respectively. We also assume that there
is no reflection of the electrons from the anode~\cite{comment-1}.

The steady state problem has a well-known solution
in the implicit form:
\begin{eqnarray} \label{v_0}
\sqrt{V_0(x) - V_i} \left(V_0 (x) + 2 V_i \right) =  3 \sqrt{\frac{2\pi e |J_0|}{\kappa_0 m}}\,x\,,\\
F_0 (x)=  - \sqrt{\frac{8 \pi m|J_0|}{ e \kappa_0}}\,\sqrt{V(x)-V_i} \,,
\end{eqnarray}
with $x$ varying from $0$ to $L$.
The current-voltage characteristic is given in a
parametric form:
\begin{eqnarray} \label{J_0}
|J_0| = \frac{\kappa_0 m}{18 \pi e L^2}
(V_f-V_i) \, (V_f+ 2 V_i)^2\,, \\
U_0=  \frac{1}{e} ({m V_f^2}/{2} - {m V_i^2}/{2})\,,
\end{eqnarray}
where $U_0$ is the applied voltage and $V_f \equiv V_0(L)$ has
the meaning of the electron velocity at the anode. At large biases,
these formulae lead to the Child's law:~\cite{virtual}
$$
|J_0| \approx  \frac{\kappa_0}{9 \pi L^2}\sqrt{\frac{2 e}{m}} U_0^{3/2}\,,\,\,
V(x) \approx {\left[ \frac{18\pi e |J_0|}{\kappa_0 m}\right]}^{1/3} x^{2/3}\,.
$$

For the time-dependent problem, we obtain two linear differential
equations:
\begin{eqnarray} \label{linearized-1}
V_0 \frac{d (V_0  V_{\omega})}{d x}
-i \omega V_0
V_{\omega} + \frac{e}{m} V_0  F_{\omega} =0 \,,\\
V_0 \frac{d F_{\omega}}{d x} -i \omega F_{\omega}
-  \frac{4 \pi |J_0|}{\kappa (\omega)\, V_0} V_{\omega} =
\frac{4 \pi}{\kappa  (\omega)} J_{\omega}\,.
\label{linearized-2}
\end{eqnarray}
An important property of these equations is the $x$-dependencies of the
coefficients via the
steady state solution $V_0 (x)$. To solve Eqs.~(\ref{linearized-1}),
(\ref{linearized-2}),  it is convenient to introduce a dimensionless
variable, $\zeta$,  in place of $x$:
\begin{eqnarray}
\nonumber\zeta =
\sqrt{\frac{V_0 (x) - V_i}{V_i}}\,,\,\,\,\,
0 \leq \zeta \leq \zeta_f \equiv \sqrt{\frac{V_f-V_i}{V_i}}\,,
\end{eqnarray}
as well as dimensionless functions in place of the electron velocity $V_{\omega}$ and the
field $F_{\omega}$:
\begin{eqnarray}
 \nu & = & \frac{\kappa(\omega) m }{36 \pi e L^2 J_{\omega}}
\frac{(V_f-V_i)(V_f+2 V_i)^2}{{V_i}^2} \,
e^{-i \theta w \zeta}\,V_0V_{\omega},\,\,\,\,\,\,\,  \label{nu} \\
 f & =&  \,\frac{\kappa (\omega) }{12 \pi L J_{\omega}}
\frac{\sqrt{V_f-V_i} (V_f+ 2 V_i)}{ \sqrt{V_i}} \,
e^{-i \theta w \zeta}\,F_{\omega}\,, \label{f}
\end{eqnarray}
 where
\begin{equation}  \label{theta}
\theta = \frac{3 \omega_{LO} L \sqrt{V_i}}{\sqrt{V_f-V_i}(V_f+ 2 V_i)}\,,\,\,\,w = \frac{\omega}{\omega_{LO}}.
\end{equation}
In such a formulation, the results are parametrically dependent on  dimensionless applied voltage, $u_0$,
and initial velocity, $v_i$,:
\begin{equation} \label{u_0}
u_0 = \frac{e~U_0}{\hbar \omega_{LO}}\,,\,\,\,v_i = \frac{V_i}{V_{LO}}\,,\,\,\,
V_{LO}=\sqrt{\frac{2 \hbar \omega_{LO}}{m}}\,.
\end{equation}
   Then, Eqs.~(\ref{linearized-1}) and (\ref{linearized-2}) can be
rewritten in a simpler form:
\begin{eqnarray}\label{system-1}
\frac{d \nu}{d \zeta} + \large[ 1 + \zeta^2\large] f =0\,,\,\,\,
\frac{d f }{d \zeta} - \frac{ I (w)}{\large[1 + \zeta^2 \large]^2}
\,\nu = e^{- i \theta w \zeta}\,,\,\,\,
\end{eqnarray}
Here, only the second equation parametrically depends on frequency via
the resonant factor $I(w) =  {2 \kappa_0}/{\kappa (w\, \omega_{LO})}$ and
the oscillating exponential in the right hand side ({\it rhs}). The latter determines
time-of-flight resonances  of ballistic electrons at a given frequency and, particularly,
the optical phonon - transit time resonance at $w \approx 1$.
According to the above discussion, the
boundary conditions for Eqs.~(\ref{system-1}) are $\nu(0) =  f (0) = 0$.
If $\nu(\zeta)$ and $f(\zeta)$ are found, coordinate dependencies $V_{\omega}(x)$ and $F_{\omega} (x)$
can be recovered by using Eqs.~(\ref{nu}), (\ref{f}) and  the following relationships:
\begin{equation}  \label{zeta-x}
\zeta(\zeta^2 +3) = 3 \sqrt{\frac{2 \pi e |J_0|}{\kappa_0 m V_i^3}}\, x \equiv \zeta_f(\zeta_f^2 +3) \,\frac{x}{L}\,.
\end{equation}
It is useful to indicate that maximum of the phase of the exponential in the {\it  rhs} of the second
Eq.~(\ref{system-1}) is:
\begin{equation} \label{om-tau}
\theta w \zeta_f = \omega \tau_{tr},
\end{equation}
where the time-of-flight, $\tau_{tr}$, is determined as
$$\tau_{tr } =\int_0^L  \frac{d x}{V_0(x)} = \frac{3 L}{(3+\zeta^2_f ) V_i}\,.$$

Eqs.~(\ref{system-1}) can be solved {\em exactly} in terms of the
Legendre functions~\cite{Legandre}, $P [a,b,x]$ and $Q[a,b,x]$.
Indeed, from the system of Eqs.~(\ref{system-1}) one can obtain
a single second order differential equation,
\begin{equation} \label{2-order}
(1+\zeta^2) \frac{d^2 \nu}{d \zeta^2} -2 \zeta \frac{d \nu}{d \zeta}
+ I(w) \nu = - (1+\zeta^2)^2 e^{-i \theta w \zeta}\,.
\end{equation}
Corresponding homogeneous equation has the following independent
solutions:
\begin{eqnarray}  \nonumber
({1+\zeta}^{2})\times P\left[ \frac{\sqrt {9-4 I(w)}-1}{2},2,i \zeta \right]\,, \\
({1+\zeta}^{2})\times Q\left[ \frac{\sqrt {9-4 I(w)}-1}{2},2,i \zeta \right]\,. \nonumber
\end{eqnarray}
Two independent solutions of  a linear homogeneous second order
differential equation allow one to construct a general solution to
nonhomogeneous Eq.~(\ref{2-order}). The resulting expression for
this solution is cumbersome and is not presented in this paper.
Instead, we shall concentrate on the discussion of conclusions, which
follow from this solution.

\begin{figure}
\includegraphics[scale=]{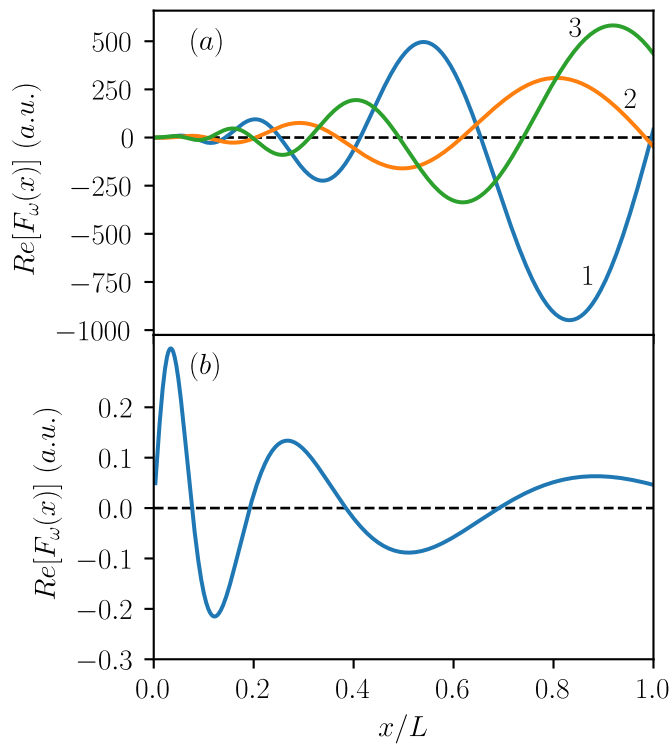}
\caption{(a): 'Snapshots' of the field-coordinate dependencies
for three frequencies close to $\omega_{LO}$: 1 - $w=0.997$, 2 - $0.993$, 3 - $w=0.9995$.
(b): The same as in (a) for the case of neglecting lattice polarization ($\kappa_{\infty}=\kappa_0$) and
$w=0.997$. Assumed parameters: $L=75$~nm, $g=0.005,~u_0=2, v_i=0.2.$
'Snapshots' are presented for the moment $t=0$.}
\label{fig-1}
\end{figure}

At the zero boundary conditions,
the time-dependent variations of the velocity, $Re [V_{\omega} e^{- i
\omega t}]$, and the field, $Re [F_{\omega} e^{-i \omega t}]$, are induced
entirely by the external {\em ac} electric bias. By solving
Eqs.~(\ref{system-1}), (\ref{2-order}) for $\nu(\zeta)$ and $f(\zeta)$, and using formulae (\ref{nu}), (\ref{f})
and (\ref{zeta-x}), we found that in the RFR   the velocity and the field are critically dependent on
the frequency. In Fig. {\ref{fig-1}} (a) we show 'snapshots' of the field-coordinate dependencies
for three frequencies close to $\omega_{LO}$ at the same stationary current (voltage) and a
given amplitude of the alternative current, $J_{\omega}$, which is set a real and positive value.
The fields are oscillating functions of $x/L$ with growing amplitudes along the electron flow.
From Eq.~(\ref{PP}) it follows that the lattice polarization, $P_{\omega} ({x}/{L})$,
behaves similarly to the alternative field, however, in the RFR it is opposite in  phase.
The effect of growing field amplitudes can be directly seen from Eq.~(\ref{system-1}), where
parameter $Re[I(w)]$ is negative in the RFR and both $Re[I(w)],\,Im[I(w)]$ become large at
$w \rightarrow 1$ (i.e., $\omega \rightarrow \omega_{LO} $).  For comparison, in Fig. {\ref{fig-1}} (b)
we present  the 'snapshot' of the field-coordinate dependency for the case, when the lattice
polarization is neglected ($\kappa_{\infty}=\kappa_0$). The assumed parameters of the diode and the
currents $J_0$ and $J_{\omega}$ are the same as for Fig. {\ref{fig-1}} (a).
It is seen, that the alternative field is still
oscillating, but its amplitude  decreases along the electron flow.

By using real and imaginary parts of $V_{\omega}$ and $F_{\omega}$ we found that spatio-temporal
dependencies of the electron velocity and the electric field are in the form of
waves propagating from the cathode to the anode; their amplitudes increase toward the anode.

\begin{figure}
\includegraphics[scale=]{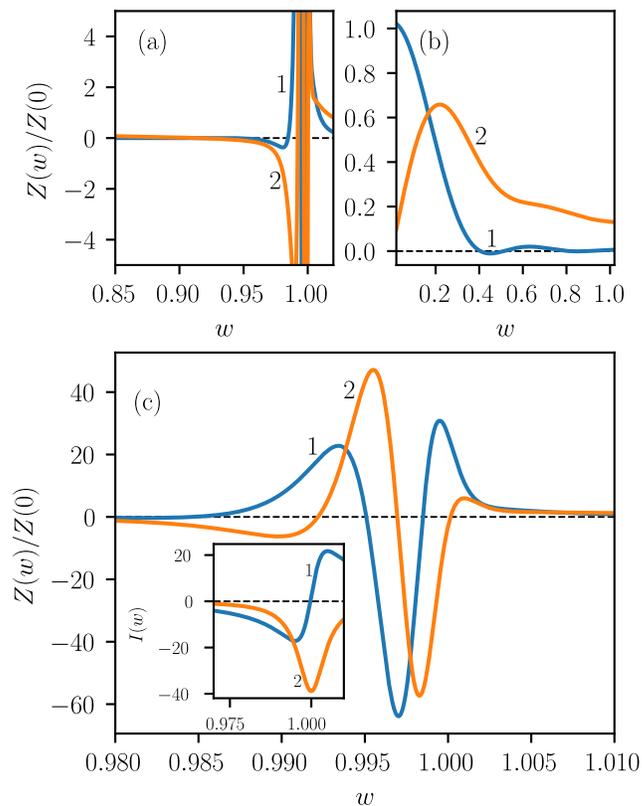}
\caption{Real and imaginary parts of
the diode impedance as  functions of the frequency.
(a):  $Z(w)$ for a magnified frequency range;
(b):  $Z(w)$ for the case $\kappa (\omega) = \kappa_0 = const$;
(c):  $Z(w)$ near $w=1$. In the inset: function $I(w)$.
Numerical parameters for calculations are the same as in Fig.~\ref{fig-1},
$Z(0) = 0.9 \times 10^{-6}$ $\Omega$~cm$^2$.}
 \label{fig-2}
\end{figure}

Once the $\nu (\zeta)$ dependence is obtained, we can calculate
the diode impedance (per unit area of the device cross-section) as
\begin{eqnarray}
\displaystyle
 Z (\omega) =  \frac{-\int_{0}^{L}  d x F_{\omega}}{J_{\omega}}  =\\
 \frac{{36 \pi L^2 } V_i^2}{ {\kappa (\omega)} (V_f-V_i)(V_f+2 V_i)^2}
 \int_0^{\zeta_f} d \zeta \frac{d \nu (\zeta)}{d \zeta} \,e^{i \theta \zeta}\,.
\nonumber
\end{eqnarray}
In Fig.~\ref{fig-2},  examples of the calculations of the diode
impedance, $Z (w)$, are shown. For this particular diode, we found
$\tau_{tr} \approx 0.28$ ps, $\omega_{LO} \tau_{tr} /2 \pi \approx 2.5$.
In the main part of the figure (panel (c)), the
real and imaginary parts of  $Z(w)$ are presented for the narrow
frequency interval near $w=1$ ($\omega \rightarrow \omega_{LO}$).
For the used parameters (see
discussion below), the RFR is $0.915 < w  < 1$. In Fig. \ref{fig-2} (a),
$Re \left[Z (w)  \right]$ is shown for the magnified frequency
interval, $0.85< w < 1.05$. In Fig.~\ref{fig-2} (b),
the same is shown for the case, when the dynamic polarization of the
lattice is neglected, i.e., $\kappa_{\infty} = \kappa_0$. For the latter case,
the frequency dependence of $Z (\omega)$ is solely due to the time-of-flight
effects (see, for example, Ref.~\onlinecite{Grib-1,JNO}).

From presented results it follows that the lattice vibrations in
polar materials give rise to strong modifications of the high-frequency
resistivity in the RFR near the optical phonon frequency.
Both $Re[Z(w)]$ and $Im[Z(w)]$ demonstrate a few oscillations with
amplitudes tens time larger than the differential resistivity in the
steady state regime, $Re[Z(0)]$. Particularly, $Im[Z(w)]$ shows
alternative inductive and capacitive characters, in opposite to the
exclusively inductive character in the case of dispersionless permittivity shown
in Fig.~\ref{fig-2}(b). While $Re[Z(w)]$ shows two narrow bands
$0.94 < w < 0.985$ and $0.995 < w < 0.999$
with negative dynamic resistance,  that reaches values $minRe[Z]_1= - 0.36\times Z(0)$ at
$w_{m1} = 0.98$ and $minRe[Z]_2= - 61.9\times Z(0)$ at $w_{m2}  = 0.997$.
The second very narrow bands is confined between two bands of
positive dynamic resistances  with sharp peaks $maxRe[Z] = 22.1 \times Z(0)$ at
 $w_{M1} = 0.993$ and $maxRe[Z] = 29.8 \times Z(0)$ at  $w_{M2} = 0.999$.
For comparison,  in the case of the dispersionless permittivity the NDR is predicted in the wide band,
$0.81<w  < 0.9$, but the amplitude of the effect is very small: $minRe[Z] \approx  - 0.001 \times Z(0)$.

The physics underlying the effects of  positive and negative  $Re[Z(w)]$
is related to correlation between the electronic motion and the self-consistent alternative field.
This is well illustrated by the field-coordinate dependencies presented in
Fig.~\ref{fig-1}. Indeed, during dynamic processes of a given frequency, $\omega$, gain/loss of local electron energy
is determined by the electrical power density averaged over the time period: $W_{\omega} (x) = \overline{J(x,t) \times F(x,t)}
 = Re[J_{\omega} F^*_{\omega}]/2$. For Fig.~\ref{fig-1} it was set $J_{\omega}$ as real and positive,
 thus for presented results $W_{\omega} (x) \propto Re[F_{\omega} (x)]$.
 One can see that for all cases the power density $W(x)$ is oscillating function on the coordinate.
For curves $2,\,3$ of Fig.~\ref{fig-1} (a)  and for Fig.~\ref{fig-1} (b) the {\it total electric power} transferred to the electrons
 (i.e., $\int W(x) d x$) is {\it positive}, which corresponds to an absorption of high-frequency electrical power in the diode base
 (corresponding resistivity $Re[Z(\omega)] >0$). However, for curve 1 of  Fig.~\ref{fig-1} (a), the  total power transferred to
 the electrons  is evidently {\it negative}, which means gain of high-frequency electric power
 (corresponding resistivity $Re[Z(\omega)] < 0$).

\begin{figure}
\includegraphics[scale=]{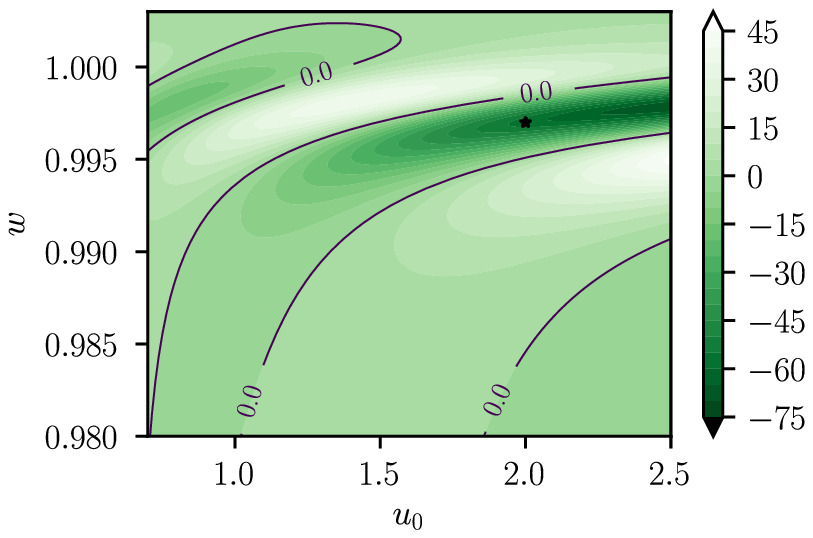}
\caption{Density plot of $Re[Z]$ in variables $ \{ w, \,u_0 \}$.
The lines ceparate regions with the NDR.  The dot indicates parameters $w$ and $u_0$,
for which  results   presented in Figs.~\ref{fig-1}, \ref{fig-2} and \ref{fig-3} are obtained.
}
 \label{fig-4}
\end{figure}

More general pattern of the resonances in the high-frequency resistivity of the short
diode is illustrated in Fig.~\ref{fig-4}, where the density plot of $Re[Z]$  is presented as
a function of two variables: $w$ and $u_0$. The thin lines mark regions with the NDR.
The narrowness of these regions is explained by the resonance character of the discussed effects.
These results demonstrate that the resonant character of the dielectric
permittivity of Eq.~(\ref{dielectric}) cardinally modifies the diode response in the RFR
and leads to regions with the NDR.

Generally, the diode response, $Z(w)$, is strongly influenced by
the two main factors entering Eqs.~(\ref{system-1}), (\ref{2-order}).
The first factor, $I (w)$, determines the coupling between the charge,
its motion and the electrostatic field. In the narrow frequency
interval near $\omega_{LO}$, this factor critically depends on
both the dimensionless frequency, $w$, and the dimensionless optical
phonon damping,
$g \equiv \gamma/\omega_{LO}$. The resonant character of the
$I(w)$-dependence
is illustrated in the Inset  of Fig.~\ref{fig-2}  (c) for
$g=0.005$ (the discussion of $\gamma$ and $g$ is presented below).
The minimum ({\it m}) and maximum ({\it M}) of $Re[I(w)]$ are realized at $w_{m,M}=\sqrt{1 \mp 2 g}$,
respectively. In the limit $g \ll 1$,  at  $w=w_m$ we find large negative values of $Re[I],\,Im[I]$:
\begin{equation} \label{min}
Re[I] \approx Im [I] \approx -  \frac{\kappa_0 -\kappa_{\infty}}{2 g \kappa_{\infty}}\,.
\end{equation}
In the  above analyzed  examples with very close frequencies
$w_{M1},\,w_{m2}, w_{M2}$  corresponding to the maxima and minimum of $Re[Z(w)]$, we obtain
sufficiently different values of $I(w)$:
$I(w_{M1}) =  -14.0 - 27.3 i$,  $I(w_{m2}) = -15.3-12 i$ and  $I(w_{M2}) =
- 1.3 - 36.7 i$. These result in distinct behavior of the high frequency
fields and impedances, as shown in Figs.~\ref{fig-1} and \ref{fig-2}.
\begin{figure}
\includegraphics[scale=]{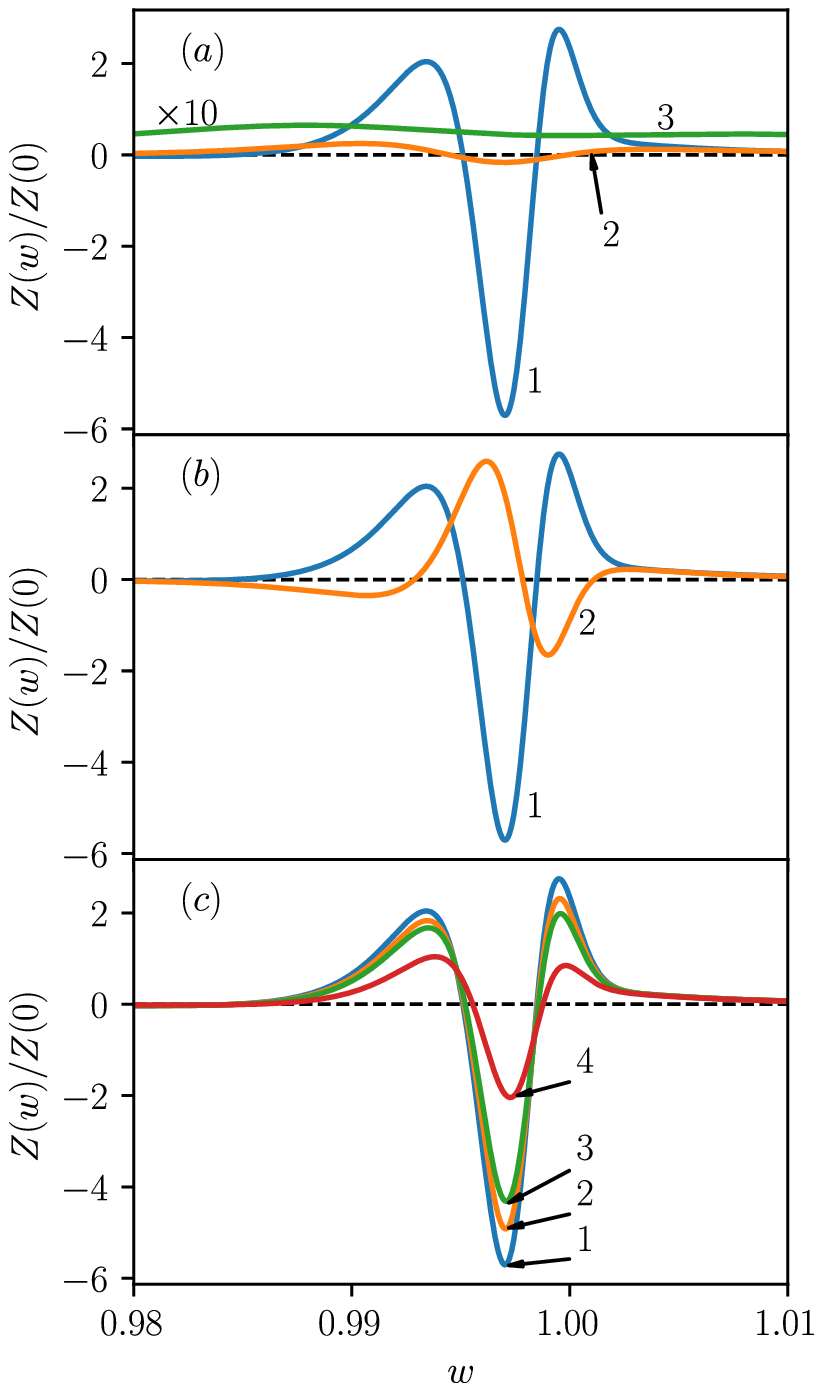}
\caption{Frequency dependencies of $Re[Z]$.
(a): 1 - $g = 0.005$, 2 - $g= 0.01$, 3 -  $g=0.02$.
(b): $g=0.005$, 1 -  $v_i = 0.2 $, 2 -  $v_i=0.3$.
(c): $g= 0.005$, 1 - $\tau_{sc} \rightarrow \infty$,
2 -  $\tau_{tr}/\tau_{sc} = 0.37$,
3 -  $\tau_{tr}/\tau_{sc} = 0.7$,
4 - $\tau_{tr}/\tau_{sc}= 1.5$.  Other
diode parameters and $Z(0)$ are as in Figs. \ref{fig-1} and \ref{fig-2}.}
 \label{fig-3}
\end{figure}
Fig.~\ref{fig-3} (a) illustrates the large effect of the phonon damping,
$\gamma~(g)$, on the resonance features of $Re[Z(\omega)]$.  At small
value of the damping, $g = 0.005$, the amplitudes of the resonant
variations of the high frequency resistivity are large. Twofold increase
in $g$ leads to more than one order decreasing in
these amplitudes. Subsequent twofold increase in $g$ totally
suppress the resonance effects.

Note, according to Eq.~(\ref{min}) for frequencies near $\omega_{LO}$,
the factor $I(w)$ is proportional to $\frac{\kappa_0-\kappa_{\infty}}{\kappa_{\infty}}$,
which characterizes a relative contribution of the polar lattice vibrations to
the permittivity, as well as the {\it electron-optical phonon coupling}.~\cite{RTD-1}
Thus, materials with larger polar properties (larger ionicity) should show
larger discussed effects.

The second factor influencing the studied resonant effects is the product  given by Eq.~(\ref{om-tau}).
Since the major effects are expected at $\omega$ close to $\omega_{LO}$,
this factor depends  mainly on the diode length, $L$, and the applied {\em dc} voltage, $U_0$, via
the $\zeta_f(U_0)$-dependence.

It is necessary to mention that in the model of the monoenergetic
electron injection the discussed results depend on the initial electron
energy (velocity, $V_i$). Varying $V_i$ at fixed other parameters,
we found that the resonance features in the resistivity are always observable,
however, there are frequency shifts in their position. Fig.~\ref{fig-3} (b)
illustrates the effect of initial velocity on the resistivity,  $Re\left[Z(\omega)\right]$.

The above results were obtained for strictly collisionless motion of the
electrons in the diode base. Approximately, the effect of electron
scattering can be estimated by introducing a "frictional force",
$- \frac{v}{\tau_{sc}}$ into the right hand side of the Newton Eq.~(\ref{N}),
with $\tau_{sc}$ being the electron scattering time. Accordingly, in
the right hand side of the first equation from (\ref{system-1}),
we need to add the term $\frac{\tau_{tr}}{\zeta_f \tau_{sc}}$.
In Fig.~\ref{fig-3} (c) we present $Re[Z(w)]$ calculated for different
ratio $\frac{\tau_{tr}}{\tau_{sc}}$ at fixed other parameters.
From these estimations it follows that though processes of electron
scattering suppress the resonant effects in the high-frequency resistivity,
the effects persist considerably large even at
$\frac{\tau_{tr}}{\tau_{sc}} \sim 1$, i.e., for quasi-ballistic electron
motion.\cite{Grib-1}

\section{Examples of nanodiodes based on III-V compounds. } \label{S-3}

In this Section we consider the resonant effects in the high-frequency
resistivity for a few particular nanodiodes based on polar III-V semiconductor
compounds.

As discussed above, the value of the optical phonon damping, $\gamma$,
is critically important for the resonant effects under consideration.
Thus we shall begin with a remark about the available data on phonon decay.
Two different approaches were used to study the optical phonon decay. The
first is incoherent Raman spectroscopy of noncoherent nonequilibrium
phonons based on time-resolved incoherent Raman
measurements,~\cite{Linde,Kash} or on measurements of the linewidth
of spontaneous Raman scattering.~\cite{Evans,Paraya} This approach provides
experimental measurements of {\it the time of relaxation of
the phonon energy}. The second approach uses the time-resolved
coherent nonlinear technique (anti-Stokes Raman scattering). In such an approach,
an optical phonon is excited coherently and the subsequent dynamics of its dephasing can
be monitored.~\cite{Vallee,Bron} As a result, {\it the phonon dephasing time}, $\tau_{ph}$,
can be extracted. Namely, the latter parameter determines the value $\gamma$.
For polar materials GaAs, InP,  dephasing processes of the optical phonons and the time, $\tau_{ph}$
 has been studied and measured in detail.~\cite{Vallee,Bron}

{\it GaAs diodes.} For numerical results presented in Figs.~\ref{fig-1} - \ref{fig-4},
we used the material parameters of GaAs:
$\kappa_0= 12.9$, $\kappa_{\infty} = 10.8$, $ \hbar\omega_{LO}
= 36.3$ meV ($\omega_{LO}/2 \pi = 8.9$ THz,
$\omega_{TO}= 0.92\,\omega_{LO}$) and $m=  0.067\,m_0$, where $m_0$
is the free electron mass.
The assumed ballistic transport regime is valid if the mean free path
of the electrons exceeds the diode length $L$. The mean free path
depends drastically on the electron energy.  In a high quality
intentionally undoped GaAs sample at the lattice temperature below $77$~ K, the
electrons with the energy less than $\hbar \omega_{LO}$ have the
mean free path up to  $1$~$\mu$m. The electrons with the energy
above $\hbar \omega_{LO}$ relax quickly through emission of
optical phonons. Direct Monte-Carlo simulation of the dynamics
of the injected electrons in GaAs at $T \leq 77$~K  proves that the fraction of
the ballistic electrons at the distance $L = 75$~nm  equals 0.74 for
$ u_0 = 2$. These values of $L$ and $U_0$ are used in
calculations presented in Figs,~1 - 3 . The relevant steady state electrical characteristics
are:  the current density $J_0 = 59$~ kA/cm$^2$ and $Z(0) = 0.9 \times 10^{-6}$~$\Omega$ cm$^2$.
For Figs.~\ref{fig-1} - \ref{fig-4} and \ref{fig-3} (a), (c), initial velocity of the injected electrons
is set $v_i = 0.2$, which corresponds to the electrons injected from a cathode at $30$~K,
while $v_i=0.3~V_0$ corresponds to the electrons injected at $T =50$~K (Fig.~\ref{fig-3} (b), curve 2).
The latter results indicate a considerable dependence of the resonances on initial velocity
of injected electrons: particularly, the frequency bands of the NDR are shifted and amplitudes of
the effects are changed.

For high quality GaAs crystals, it was found the optical phonon dephasing times:
$\tau_{ph} \approx 2$~ps at $T=300$~K, $6.4$~ps at $T=77$~K, $9$~ps at $6$~K.~\cite{Vallee}
Corresponding phonon damping parameter is $g = 0.005...0.001$,  In the model calculations presented
in Figs.~\ref{fig-1} - \ref{fig-4} we used $g = 0.005$.

For comparison ballistic and quasi-ballistic diodes, which is shown in Fig.~\ref{fig-3} (c),
we used the scattering times, $\tau_{sc}$, corresponding to following values of
the mobility: $2 \times 10^4,~10^4,~5 \times 10^3$~cm$^2$/(Vs), curves $2,~3,~4$, respectively.

We performed also the calculations for the GaAs diode with shorter base, $L= 50$~nm, for which
the fraction of the ballistic electrons exceeds $0.95$ and $0.8$ at $u_0 =1$ and $2$, respectively.
For such a diode, at $u_0 = 1$ we found $J_0 = 53.4$~kA/cm$^2$, the windows with the NDR,
$0.996 < w < 1$, and $minRe[Z] = -15.2 \times 10^{-6}$~$\Omega$ cm$^2$ at $w=0.998$. At $u_0=2$,
we obtained $J_0 = 132$~kA/cm$^2$ and $minRe[Z]= - 55 \times 10^{-6}$~$\Omega$cm$^2$ at $w=0.997$.

{\it InP diodes.}  In comparison with GaAs, InP crystals are characterized larger ionicity (see Eq.~\ref{min})
and larger optical phonon lifetime.  For calculations we used the following parameters of InP:
$\kappa_0= 12.5$, $\kappa_{\infty} = 9.6$, $\hbar \omega_{LO} = 43$~meV ($\omega_{LO}/2 \pi = 10.4$~ THz,
$\omega_{TO}= 0.88\,\omega_{LO}$) and $m=  0.08\,m_0$.  In high-quality InP crystals at low temperatures ($T \leq 100$~K)
the electron mobility can reach $10^5$~cm$^2$/Vs, with the mean free path up to $1$ $\mu$m. However,
 the electrons with energies $E $ above $\hbar \omega_{LO}$ rapidly emit the optical phonons,
 (this is actual for $U_0 > \hbar  \omega_{LO}/e$).  We found that  the fractions of the ballistic
electrons at the distance $L = 50$~nm are greater  than 0.95 and 0.6 for $ u_0 = 1$ and $2 $,
respectively.  At the distance $L= 75$~nm and $u_0=1$, the fraction of the ballistic
electrons is $0.9$. Thus, for these parameters the electron transport can be considered as quasi-ballistic.

InP materials are characterized by relatively large optical phonon lifetime:
$\tau_{ph}= 7.6$~ps  at $T=100$~K ($g = 0.001$), $22$~ps  at $T=77$~K ($g \approx 0.0004$) and
$40$~ps  at $T=6$~K ($g \approx 0.0002$).\cite{Vallee,gamma-1}
In the model calculations presented below we used $g = 0.002$.

We found that nanoscale InP-diodes show all discussed above resonant features of the resistivity near
the optical phonon frequency. In the case of the base length $75$~nm at $u_{0}=1$ we found
$\tau_{tr} = 0.36$~ps, $\omega_{LO} \tau_{tr}/ 2 \pi \approx 3.8$, $J_0 = 27$~kA/cm$^2$, $Z(0) = 1.8 \times 10^{-6}$~$\Omega$ cm$^2$.
There is one relatively wide frequency interval with NDR, $0.993< w <0.997$, where
$minRe[Z] = 14 \times 10^{-3}$~$\Omega$ cm$^2$  at $w=0.995$.

  In the case of the base length $50$~nm at $u_0=1$
  we found $\tau_{tr} = 0.24$~ps, $\omega_{LO} \tau_{tr} / 2 \pi \approx 2.5$, $J_0=61$~kA/cm$^2$,
  $Z(0) = 0.55 \times 10^{-6}$~$\Omega$ cm$^2$. There is one frequency interval with NDR,
  $0.993< w <0.997$, with   $minRe[Z] = -12 \times 10^{-3}$~$\Omega$ cm$^2$   at $w=0.996$.
In general, the analyzed resonant effects are considerable larger for nanoscale InP-based diodes
in comparison with the GaAs diodes.

Thus, the model of the monoenergetic electron injection predicts strong resonant
effects in high-frequency resistivity and opening frequency windows with large amplitudes of the  NDR.
Finalizing this analysis we mention that this type of the injection can be realized, for example,
by the use of the resonant tunneling through a double barrier heterostructure.~\cite{RTD-1}
Indeed, at the resonant  tunneling the spread of the energy of injected electrons can be
less or order of $0.1...1$~meV, while the current densities reach
tens to hundred of kA/cm$^2$.~\cite{RTD-1}

For more common $n^+-i$ contacts, the injected electrons are spread over
energy according to the temperature of the cathode.
Different groups of the electrons are injected with different velocities,
are characterized by different times of flight and contribute to the
resistivity resonances at slightly different frequencies.
Consequently, the overall resonance effects decrease.

To take into account the finite temperature
of the injected electrons we should apply more adequate model
for the ballistic electron transport in the nanodiodes.

\begin{figure}
\includegraphics{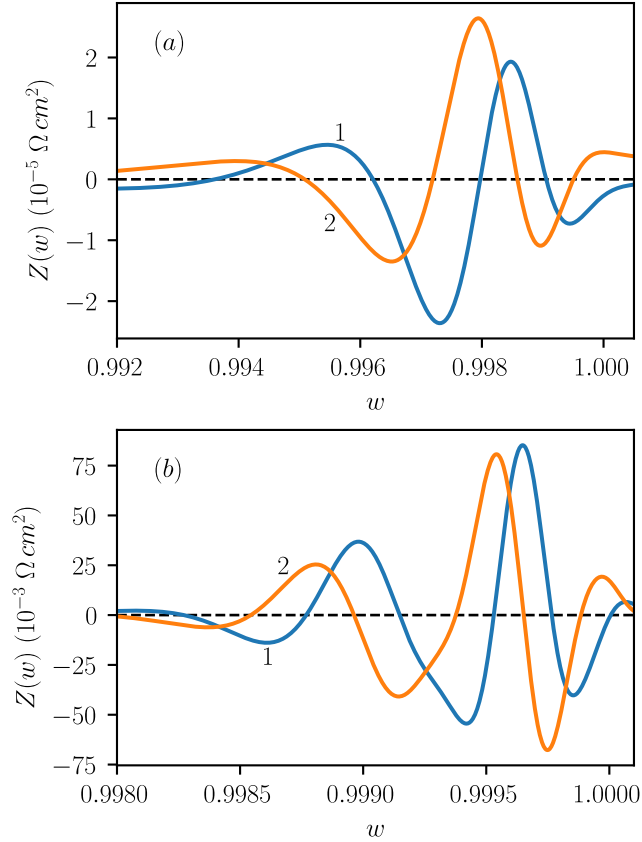}
\caption{Frequency dependencies of $Z(w)$ for the nanoscale diodes made of particular materials.
(a): the GaAs-diode at the cathode temperature $T = 50$~K.
(b): the InP-diode at the cathode temperature $T = 77$~K.
1 - $Re[Z(w)]$, 2 - $Im[Z(w)]$.
The parameters of the diodes are given in the text.}
 \label{fig-5}
\end{figure}

\section{Model based on Boltzmann transport equation for
injected ballistic electrons.}

To take into account the velocity distribution  of the injected
electrons, Eqs.~(\ref{N}) and (3) should be replaced by the
Boltzmann equation for the distribution function $\Phi (x, {\vec
v}, t)$. In the case of the ballistic electrons, the latter
equation reads:
\begin{equation}  \label{BTE}\frac{\partial \Phi}{\partial t} +
V_x \frac{\partial \Phi}{\partial x}- \frac{e}{m} F\, \frac{\partial
\Phi}{\partial V_x} = 0\,.
\end{equation}
Now two functions, $\Phi (x, {\vec v}, t)$ and $ F (x, t)$,
describe completely the system under consideration. The electron
concentration and the current can be calculated by using $\Phi
(x,{\vec v}, t)$: i.e., $n =  \int \Phi \,d^3 v,\,j =  -e \int V_x
\Phi\,d^3v$. The latter relationships and Eqs.~(\ref{dielectric}),
(4), (\ref{BTE}) comprise the necessary set of the equations. The
boundary condition to Eq.~(\ref{BTE}) is determined by the
equilibrium distribution of the incoming electrons at a given
temperature $T$ of the cathode. As described earlier, we separate
the steady state and time-dependent problems: $\Phi (x, {\vec v},
t) =  \Phi_0 (x, {\vec v}) +\Phi_{\omega} (x, {\vec v}) exp(-i
\omega t)$ and $F=  F_0 (x) +F_{\omega} (x) \exp (- i \omega t)$.
While the steady state solutions $\Phi_0 (x)$ and $F_0 (x)$ can be
found analytically (see, for example, Ref.~\onlinecite{Bulashenko}), the time dependent solutions require
the use of numerical calculations, as discussed in Ref.~\onlinecite{JNO}.

The subsequent computations confirmed the basic conclusions discussed
in Sections \ref{S-2} and \ref{S-3} for the simple model.
We also checked that the results based on the Boltzmann transport
coincide with those obtained in the simple model, when
the cathode temperature $T \rightarrow 0$.
At a finite temperature, $T$, we found that current-voltage characteristics
and $Z(0)$ are almost the same as in the simplest model. However the resonance effects
are weakened, as expected.

In Fig.~\ref{fig-5} (a),  we present $Z(w)$ for the $75$~nm GaAs diode at the cathode temperature
$T=50$~K assuming $U_0 = 73$~meV~($u_0=2$) and restricting ourselves to the case of a small dephasing optical phonon lifetime,
$\tau_{ph} = 2$~ps.  We can see that there are two frequency bands with the pronounced  resonances and the  NDR:
$0.995<w<0.997$
with $minRe[Z]_2 = -1.3 \times 10^{-5}$~$\Omega$ cm$^2$ at $w=0.996$ and $0.9985 < w< 0.9995$ with
$minRe[Z]_3 = - 1.1 \times 10^{-5}$~$\Omega$ cm$^2$ at $w=0.999$.
Further increase in $T$ suppresses the resonant effects, for example, at $T=77$~K we found only one
frequency band with the NDR: $0.996< w < 0.998$ and $minRe[Z]= - 0.7 \times 10^{-5}$~$\Omega$ cm$^2$
at $w = 0.997$.

Similar results we obtained for the $50$~nm GaAs diodes at $T=50$~K and $ 77$~K under the voltage
$U_0 \approx 36...80$~meV.  For example, at $T=50$~K and $U_0=73$~meV we found
$minRe[Z] = 1.8 \times 10^{-5}$~$\Omega$ cm$^2$.

In Fig.~\ref{fig-5} (b), calculations of $Z(w)$ for the $75$~nm InP diode are shown at the cathode temperature
$T=77$~K and $U_0 = 43$~meV($u_{0} = 1$) and the dephasing optical phonon lifetime $\tau_{ph} = 7.6$~ps.
We found three very narrow frequency bands with the NDR near $w=1$: $0.9983 < w < 0.9988$ with
$minRe[Z]_1 = -2.4 \times 10^{-3}$~$\Omega$ cm$^2$ at $w=0.9988$; $0.9991<w<0.9994$ with $minRe[Z]_2
=- 14 \times 10^{-3}$~$\Omega$ cm$^2$ at $w=0.9993$; $0.9996 < w < 0.9999 $ with $minRe[Z]_3 = - 10.6 \times
10^{-3}$~$\Omega$ cm$^2$ at $w=0.9998$.
We found that the resonant effects under consideration remain at elevated temperatures.
For example, at $T=150$~K and $u_0=1$ two most pronounced frequency bands with the NDR
were obtained as $0.9989 <w<0.9994$ with $minRe[Z] = 3.6 \times 10^{-4}$~$\Omega$ cm$^2$ at $w=0.9991$,
and $0.9996<w<0.9999$ with $minRe[z] = 4.3 \times 10^{-4}$~$\Omega$ cm$^2$.

Note, for both GaAs and InP diodes we used small dephasing optical phonon times ($\tau_{ph} =2$~ps and $7.6$~ps,
respectively). In high quality materials  $\tau_{ph}$ can be two to three times larger, thus
the amplitude of the NDR can be increased by one order of value.

\section{Discussion and summary}

We investigated ultra-high frequency electron response of nanoscale ballistic
diodes made by polar materials. For such nanoscale diodes the polarization lattice
vibrations provide the dynamic screening and significantly contribute to the high
frequency response of the devices.  When characteristic frequencies of ballistic
electron transfer across the diode are of the order of the polarization lattice vibrations,
we found large resonant effects in the frequency dependent resistivity and an enhanced
negative dynamic resistivity in the  reststrahlen frequency range.

In this frequency range, as the result of the specific dynamic screening, the alternative self-consistent
electric field, the lattice polarization and the electron (average) velocity are oscillating functions
of the coordinate with  growing amplitudes along the electron flow.
While outside of the RFR amplitudes of these characteristics decrease
along the flow. In close proximity to the optical phonon frequency
both real and imaginary parts of the impedance $Z(\omega)$ show
a few oscillations with amplitudes tens time larger than the differential resistivity in the
steady state regime, $Re[Z(0)]$. Particularly, $Im[Z(\omega)]$ shows
alternative inductive and capacitive characters, in opposite to the
exclusively inductive character in the case, when the dynamic lattice polarization is neglected
(see Fig.~\ref{fig-1} (b)).
While $Re[Z(\omega)]$ becomes negative, generally, in several very narrow frequency bands
(of order of $0.2$~cm$^{-1}$~(InP) ... $0.6$~cm$^{-1}$~(GaAs) in conventional units).

The amplitudes and sharpness of the resonances are determined by the factor
(\ref{min}), which is proportional to  both a relative contribution of the polar lattice vibrations to
the permittivity and the optical phonon lifetime.  Comparison of the diodes made of
GaAs (the material with modest polar properties and a small optical phonon lifetime) and InP
(the material with larger ionicity and larger phonon lifetime) demonstrates that
the latter diodes should show considerably stronger resonant effects in the dynamic
resistivity.

It was ascertained that in the ballistic diodes the studied effects are dependent on
the temperature of the electrons injected from the cathode: an increase in the cathode
temperature suppresses the resonant effects. Meanwhile,  the resonant effects, including
the negative resistance in the RFR,  are still pronounced at $T \geq 77$~K for the GaAs
diodes and at $T \geq 150$~K  for the InP diodes.

The studied NDR of the nanoscale diodes can be utilized to amplify  and generate
far-infrared emission at the expanse of energy of the current. Indeed, absolute values
of found specific negative dynamic resistance,  above $10^{-5}$~$\Omega$ cm$^2$ for the GaAs diodes
and above $10^{-3}$~$\Omega$ cm$^2$ for the InP diodes, are well above the specific contact
resistances $Z_c$ for these materials ($Z_c \approx 10^{-6}$~$\Omega$ cm$^2$).~\cite{contacts}
Thus the necessary condition of negative resistance of whole device~\cite{TWT-1} with contacts
can be met. The current densities are estimated to be of the order of tens of kA/cm$^2$.
 Then, gain and threshold of electromagnetic generation are determined by the active
part of the diode admittance (conductance), $Re[Y(\omega)]$ with $Y(\omega)=1/(Z_c+Z(\omega))$.
The key parameter of a device is its quality factor $Q = - {Im[Y]}/{Re[Y]}$. Value $Q$ nearly
 $-1$ is favorable for easy matching to external circuits.~\cite{TWT-1,Grib-1}
As seen from Figs.~\ref{fig-5} in every spectral band with the NDR
 there is a frequency for which the latter requirement is met for both GaAs and InP diodes.
Absolute values of $Re[Y]$ are sufficient to provide an electrical
means to amplify and generate electromagnetic radiation in far-infrared spectral range.

The ultra-high frequency resonances and NDR also can be observed in
{\it planar} nanoscale diodes made from hybrid structures - non-polar
semiconducting films or two-dimensional crystals on polar substrates.
Indeed, in such structures time- and space-dependent planar electron
transport induces an alternative electric fields, which,
in turn, polarize substrate near the interface.
As a result, one can expect resonant interaction of electron flux with polarized
substrate near surface optical phonon frequencies.
Examples are graphene on polar substrates ($SiO_2,\,SiC,\,hBN,\,HfO_2$, etc.).
Different aspects of interaction of graphene electrons with polar
substrates were studied in numerous papers (see, for example,~\cite{int-1,int-2,int-3,int-4}).
High mobilities in these structures facilitate ballistic electron transport over
large length scales ($\sim$hundreds $nm$)~\cite{graph-1,graph-2,graph-3}.
These properties are favorable for observation of ultra-high frequency
resonances in resistivity of planar nanoscale diodes based on
the hybrid structures, however discussed effects require additional analysis.

In summary, nanoscale diodes made of a polar
material with ballistic electron transport can exhibit a large
effect of optical lattice vibrations on the ultra-high frequency
electrical properties of the diodes. This results in the modification of the
time-of-flight effects and dramatic reconstruction of the
$\omega$-dependent impedence/admittance in the reststrahlen frequency range.
Particularly, narrow resonances in the diode impedance with the
negative real part are induced near the optical phonon frequency.
The resonant effects in the dynamic resistance of nanoscale GaAs and InP
diodes are studied in detail.
The obtained magnitudes of the NDR effect indicate that the
nanoscale diodes are capable to amplify and generate electromagnetic
radiation in the 10-THz range.

\begin{acknowledgments}
The authors are grateful to Dr. G. I. Singayivska for estimations of the electron ballisticity
under high applied voltage. This work was supported by Science \& Technology Center in Ukraine 
(STCU-NASU TRI project \#3922 "{\it New Technologies of THz and sub-THz Generation with Nanosize Semiconductor Heterostructures}")
\end{acknowledgments}


\begin{references}
\bibitem{comment-1}
Anode contacts absorbing fast ballistic electrons
without backscattering are discussed in
Refs.~[\onlinecite{Grib-1}] in detail.
\end{references}
\end{document}